\documentclass[slac_one]{revtex4}
\usepackage{graphicx}
\usepackage{fancyhdr}
\def\cpkkd{\rm{kg^{-1}keV^{-1}day^{-1}}}
\def\mwimp{\rm{m_{\chi}}}
\def\csnospin{\rm{\sigma_{\chi N}^{SI}}}
\def\csspin{\rm{\sigma_{\chi n}^{SD} }}

\def\sasix{\rm{SA_{6}}}
\def\sa12{\rm{SA_{12}}}

\begin{document}

\title{Limits on Low-Mass WIMP Dark Matter with an Ultra-Low-Energy Germanium Detector at 220~eV Threshold} 

\author{Shin-Ted Lin,~H. T. Wong ~~(on behalf of the TEXONO Collaboration)}
\affiliation{Institute of Physics, Academia Sinica, Taipei 115, Taiwan}

\begin{abstract}
An energy threshold of (220$\pm$10) eV was achieved at an efficiency of 50\%
 with a four-channel ultra-low-energy germanium detector
 each with an active mass of 5~g\cite{wimppaper}. This
 provides a unique probe to WIMP
 dark matter with mass below 10~GeV. With low background data taken at
 the Kuo-Sheng Laboratory, constraints on WIMPs in the galactic
 halo were derived. Both spin-independent WIMP-nucleon and  spin-dependent
 WIMP-neutron bounds improve over previous
 results for WIMP mass between 3$-$6 ~GeV.
 These results, together with those on spin-dependent couplings, will
 be presented. Sensitivities for
 full-scale experiments were projected. This
 detector technique makes the unexplored sub-keV
 energy window accessible for new neutrino and
 dark matter experiments.
\end{abstract}

\maketitle

\thispagestyle{fancy}

\section{MOTIVATIONS}
Weakly Interacting Massive Particles 
(WIMP, denoted by $\chi$) are
the leading candidates for Cold Dark Matter (CDM)\cite{pdgcdm}.
There are intense experimental efforts
to look for WIMPs through direct detection 
of nuclear recoils  in
$\chi$N$\rightarrow$$\chi$N 
elastic scattering or in the studies of the possible
products through $\chi \bar{\chi}$ annihilations.
Supersymmetric(SUSY) particles 
are the leading WIMP candidates.
The popular SUSY models prefer WIMP mass($\mwimp$) in the range
of $\sim$100~GeV, though light neutralinos 
remain a possibility\cite{lightsusy}.
Most experimental programs
optimize their design in the high-mass region
and exhibit diminishing sensitivities for $\rm{\mwimp < 10 ~GeV}$,
where an allowed region 
due to the annual modulation data
of the DAMA experiment\cite{damaallowed}
 further reinforced by the first 
DAMA/LIBRA results\cite{damalibra}  
remains unprobed. 
Simple extensions of the Standard Model with a 
singlet scalar favors light WIMPs.
Detectors with sub-keV threshold are needed 
for probing this low-mass region
and studying WIMPs bound in the solar system,
and non-pointlike SUSY candidates like Q-balls. 
This presents a formidable
challenge to both detector technology and background control.
We report the first results in an attempt towards such goals\cite{wimppaper}.

\section{EXPERIMENTAL SET-UP}

Ultra-low-energy germanium detectors(ULEGe)
is a matured technique for 
sub-keV soft X-rays measurements.  
Compared with $\rm{Al_2 O_3}$ which also has set limits\cite{cresst1}
at the sub-keV threshold ,
Ge provides enhancement in 
$\chi$N spin-independent couplings($\csnospin$)
due to the $\rm{ A^2 }$ dependence\cite{pdgcdm,cdmmaths}, 
where A is the mass number of the target isotopes.
In addition, 
The isotope $^{73}$Ge (natural isotopic
abundance of 7.73\%)  
comprises an unpaired neutron such that
it can provide additional probe to 
the spin-dependent couplings of WIMPs with 
the neutrons($\csspin$).
The nuclear recoils from $\chi$N interactions
in ULEGe
only give rise to $\sim$20\% of the observable ionizations
compared with electron recoils at the same energy.
For clarity, all ULEGe measurements discussed
hereafter in this article are 
electron-equivalent-energy,
unless otherwise stated. 

\begin{figure*}[t]
\begin{minipage}{18pc}
\includegraphics[width=80mm]{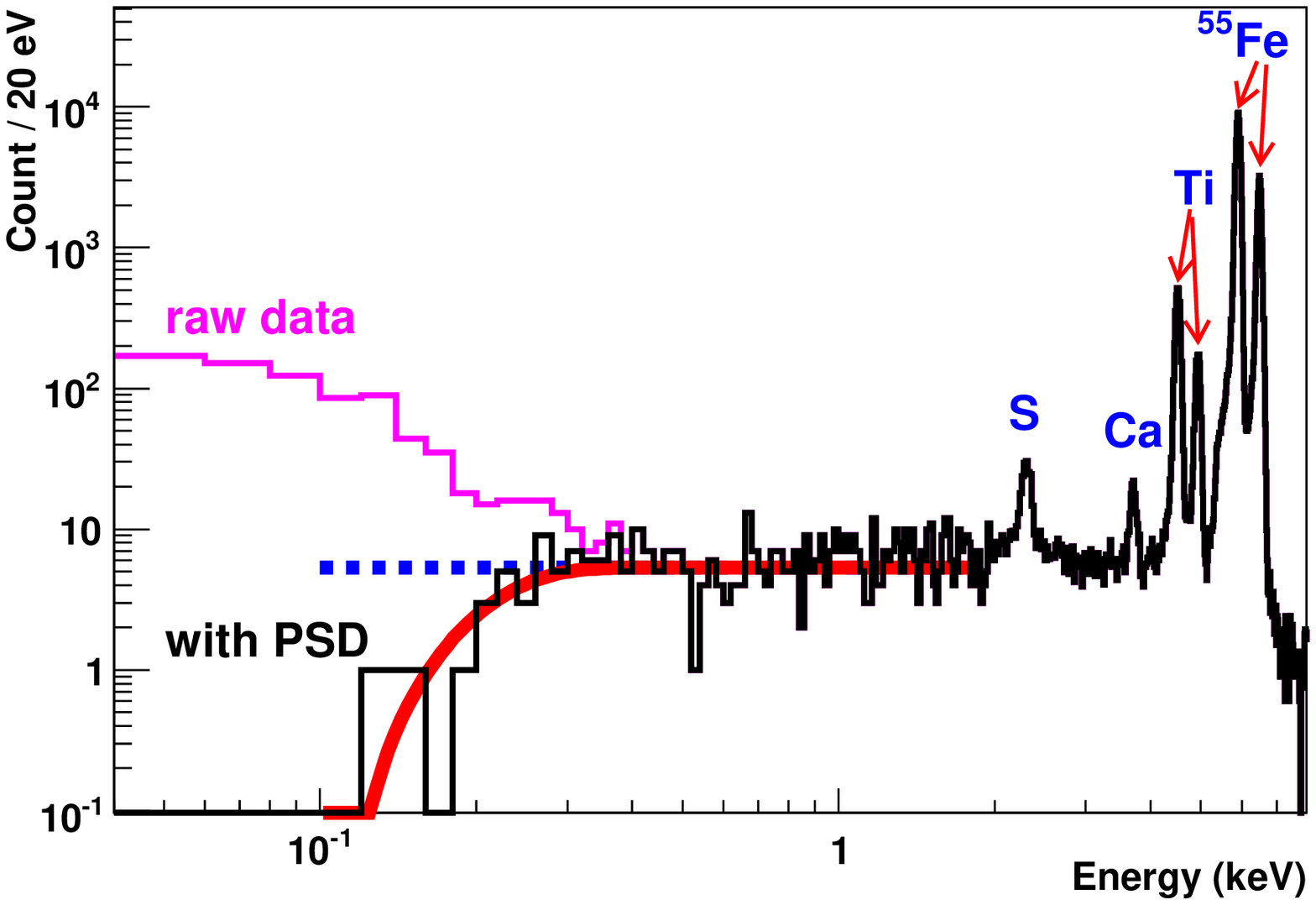}
\caption{
Measured energy spectrum of the ULEGe
with $^{55}$Fe
source together with X-ray from 
Ti, Ca and S.
The black histogram represents events 
selected by PSD cuts.
}
\label{fe55}
\end{minipage}\hspace{2pc}%
\begin{minipage}{18pc}
\includegraphics[width=80mm]{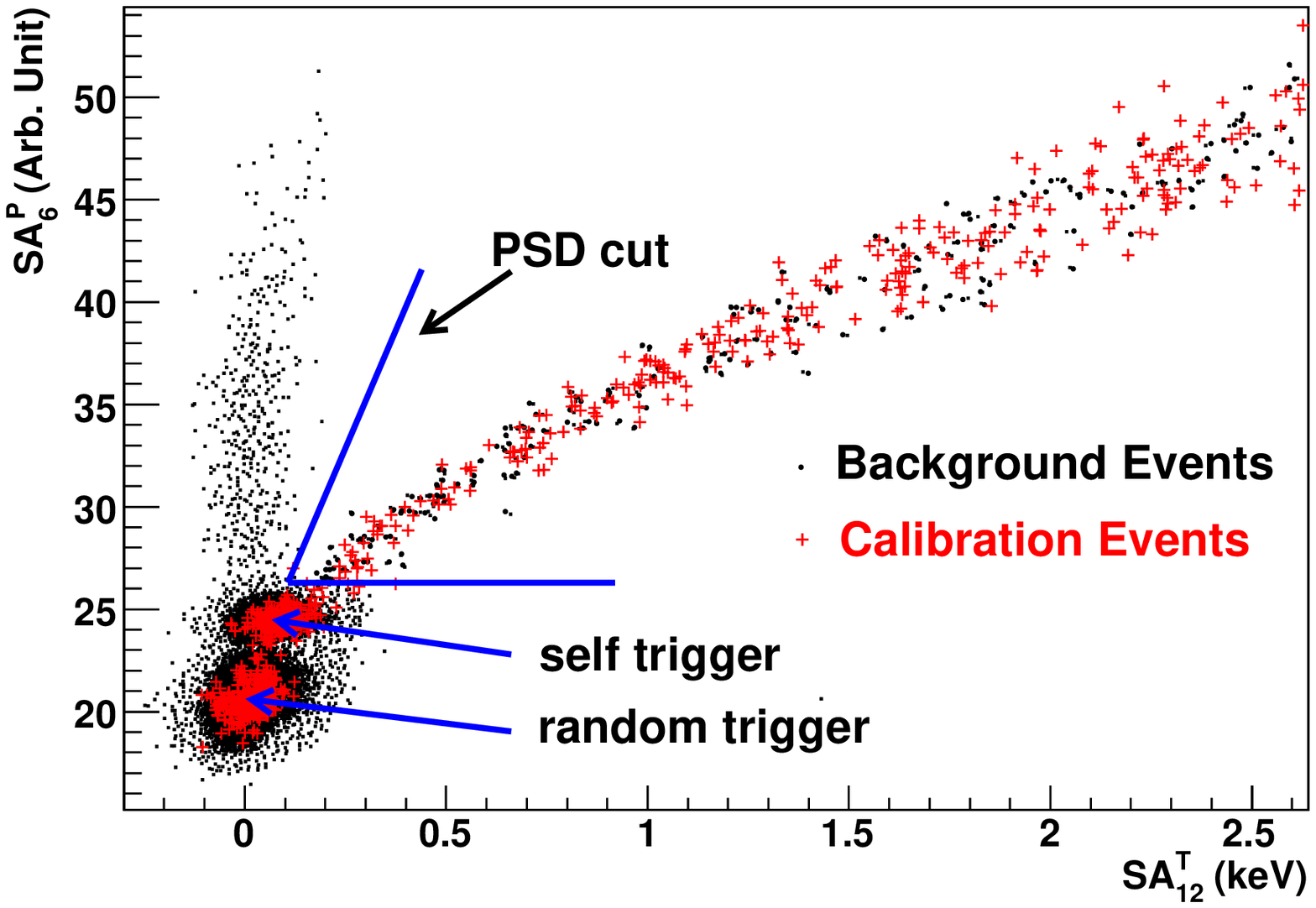}
\caption{
Scattered plots 
of the $\rm{SA^{P}_{6}}$ versus
$\rm{SA^T_{12}}$ signals,
for both calibration and physics events\cite{wimppaper}.
The PSD selection is shown.
}
\label{psd_2d}
\end{minipage}
\end{figure*}

The ULEGe array consists of 
4-element 
each having an active mass of 5~g\cite{ulege}. 
Standard ultra-low-background
specifications were adopted in its
construction and
choice of materials.
The low background measurement
was performed at 
the Kuo-Sheng(KS) Laboratory\cite{texonoprogram}
with an overburden of about 30~meter-water-equivalence.
A background level  of
${\rm \sim 1 ~ event ~ \cpkkd  }$(cpkkd) at 20~keV,
comparable with those of underground CDM experiments,
was achieved in a previous experiment with a
1-kg HPGe detector\cite{texonomagmom} for the studies
of neutrino magnetic moments\cite{munureview}.
Details of the detector and shielding components
can be refereed to Ref.~\cite{wimppaper}.

Energy calibration was achieved by the
external $^{55}$Fe sources(5.90 and  6.49~keV)
together with X-rays from
Ti(4.51 and 4.93~keV), Ca(3.69~keV),
and S(2.31~keV) in Figure~\ref{fe55}.
The RT-events provided the calibration
point at zero-energy.
The RMS resolutions for the 
RT-events and $^{55}$Fe peaks
were about 55~eV and 78~eV, respectively.

\section{SELECTION CUTS AND EFFICIENCY EVALUATIONS}

The ULEGe signals were distributed to two spectroscopy
amplifiers at 6~$\mu$s($\sasix$) 
and 12~$\mu$s($\sa12$) shaping times
and with different amplification factors.
Pulse shape discrimination (PSD) software\cite{wimppaper} 
was devised to differentiate physics events 
from those due to electronic noise,
exploiting the correlations in
both the energy and timing
information of the $\sasix$ and $\sa12$ signals
in Figure~\ref{psd_2d}.
Calibration events and
those from physics background 
were overlaid, indicating 
uniform response.

Events selected by PSD but with  Anti-Compton Veto (ACV) 
and Cosmic-Ray Veto (CRV) tags 
were subsequently rejected.
The surviving events were ULEGe signals uncorrelated 
with other detector systems
and could be WIMP candidates.
The data set adopted for the WIMP analysis
has a DAQ live time of 0.338~kg-day.
The DAQ dead time is 11\%.
The CRV and ACV selection efficiencies
of, respectively, 91.4\% and 98.3\%   
were accurately measured using
RT-events\cite{texonomagmom}.

\begin{figure*}[t]
\begin{minipage}{18pc}
\includegraphics[width=80mm]{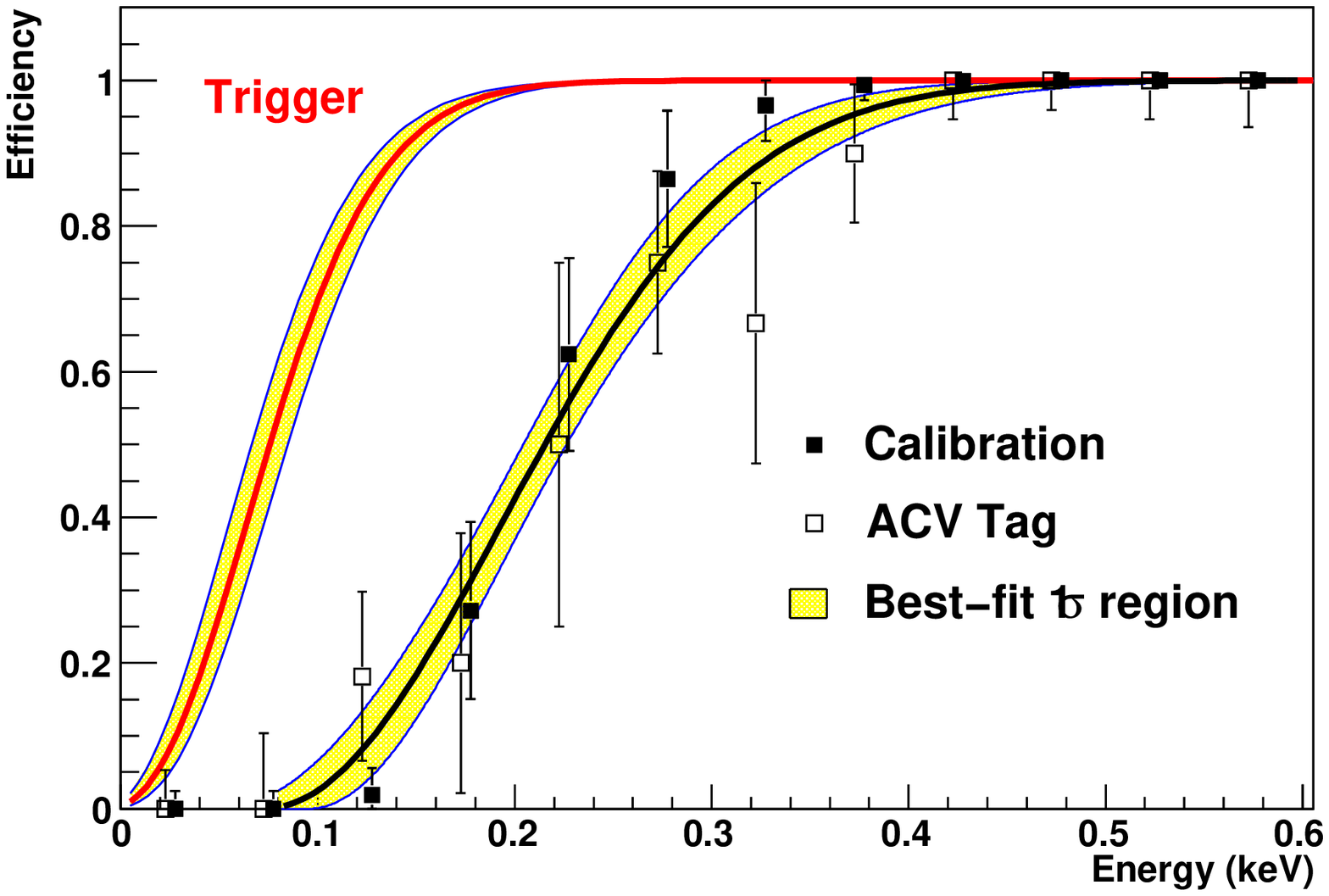}
\caption{
Trigger efficiency 
for physics events recorded by the DAQ
system and the best-fit 1$\sigma$ region
of selection efficiencies of the PSD cut, as
derived from the $^{55}$Fe-calibration
and {\it in situ} data with ACV tags.
}
\label{eff}
\end{minipage}\hspace{2pc}%
\begin{minipage}{18pc}
\includegraphics[width=80mm]{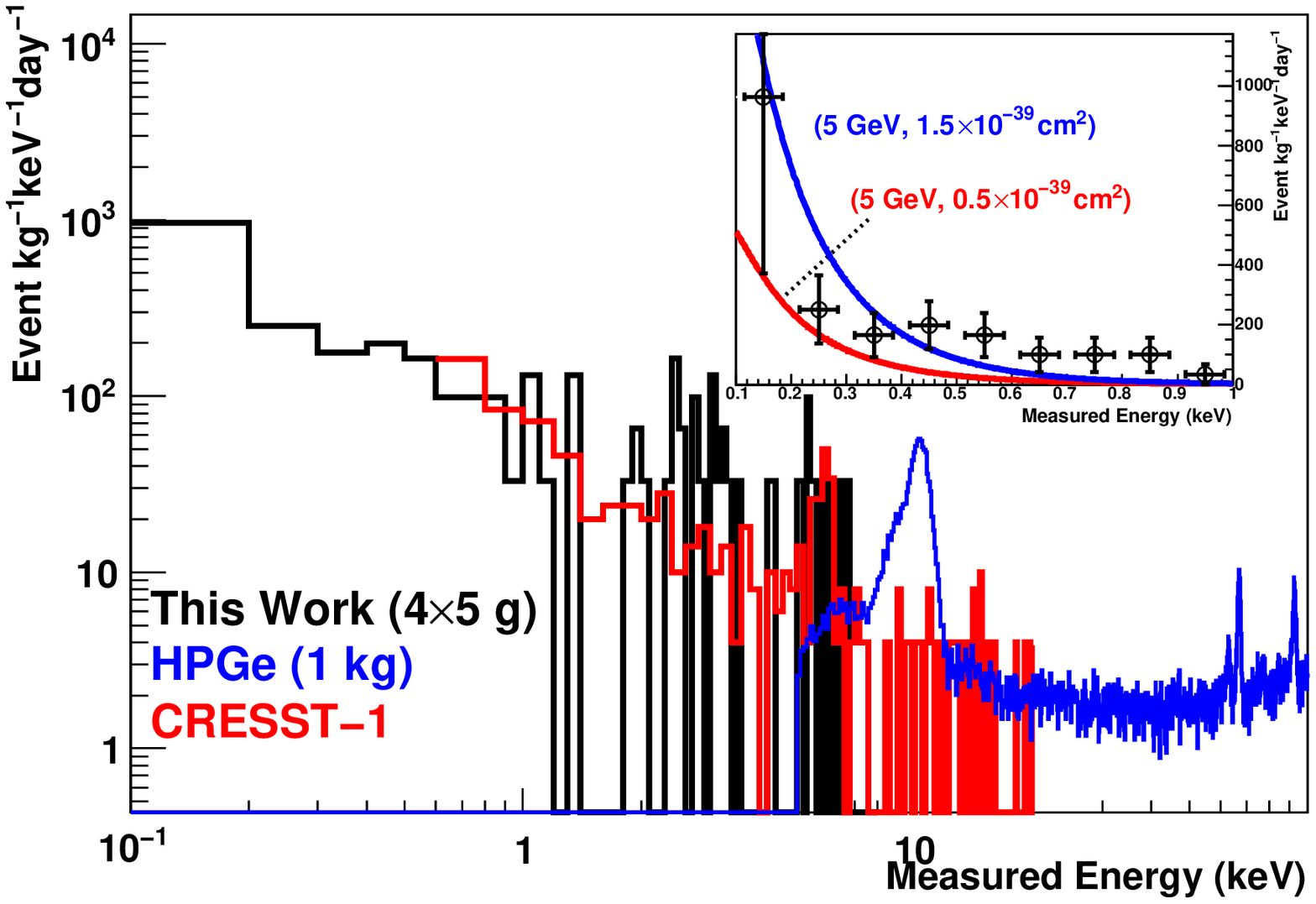}
\caption{
The measured spectrum of ULEGe with
0.338~kg-day of data,
after CRV, ACV and PSD selections. Background 
spectra of the CRESST-I experiment\cite{cresst1}
and the HPGe\cite{texonomagmom}
are overlaid for comparison. The
expected nuclear recoil spectra
for two cases
of $\rm{ ( \mwimp , \csnospin ) }$
are superimposed onto the spectrum
shown in linear scales in the inset.
}
\label{bkgspect}
\end{minipage}
\end{figure*}

The trigger efficiencies
depicted in Figure~\ref{eff}
correspond to
the fractions of the distributions
above the discriminator threshold level. 
Events in coincidence with ACV-tags 
are mostly physics-induced.
The fraction of these events surviving 
the PSD cuts represents
the PSD efficiency.
Alternatively,
the deviations of the PSD-selected events
from a flat distribution in the
low energy portion of 
$^{55}$Fe calibration spectrum of Figure~\ref{fe55} 
provided the second measurement.
Consistent results were obtained with both approaches,
as depicted in Figure~\ref{eff}.
The efficiencies and their uncertainties
adopted for analysis were
derived from a best-fit on the combined data set.
A threshold of (220$\pm$10)~eV  was achieved with
a PSD efficiency of 50\%.

\section{RESULTS}

The ULEGe spectrum normalized in cpkkd unit
after the CRV, ACV and PSD selections
is displayed in Figure~\ref{bkgspect},
showing comparable background
as CRESST-I\cite{cresst1}.
The formalisms followed those of Ref.~\cite{cdmmaths}
using standard nuclear form factors,
a galactic rotational velocity of
${\rm 230 ~ km \, s^{-1}}$ and 
a local WIMP density of $\rm{0.3 ~ GeV \, cm^{-3}}$ 
with Maxwellian velocity distribution.
The unbinned optimal interval method 
as formulated in Ref.~\cite{yellin}
and widely used by current CDM experiments
was adopted to derive the upper
limits for the possible $\chi$N event rates.
Corrections due to QF, detector resolution
and various efficiency factors 
were incorporated.
The energy dependence of QF in Ge was
evaluated with the TRIM software package\cite{trim}.

\begin{figure*}[t]
\begin{minipage}{18pc}
\includegraphics[width=80mm]{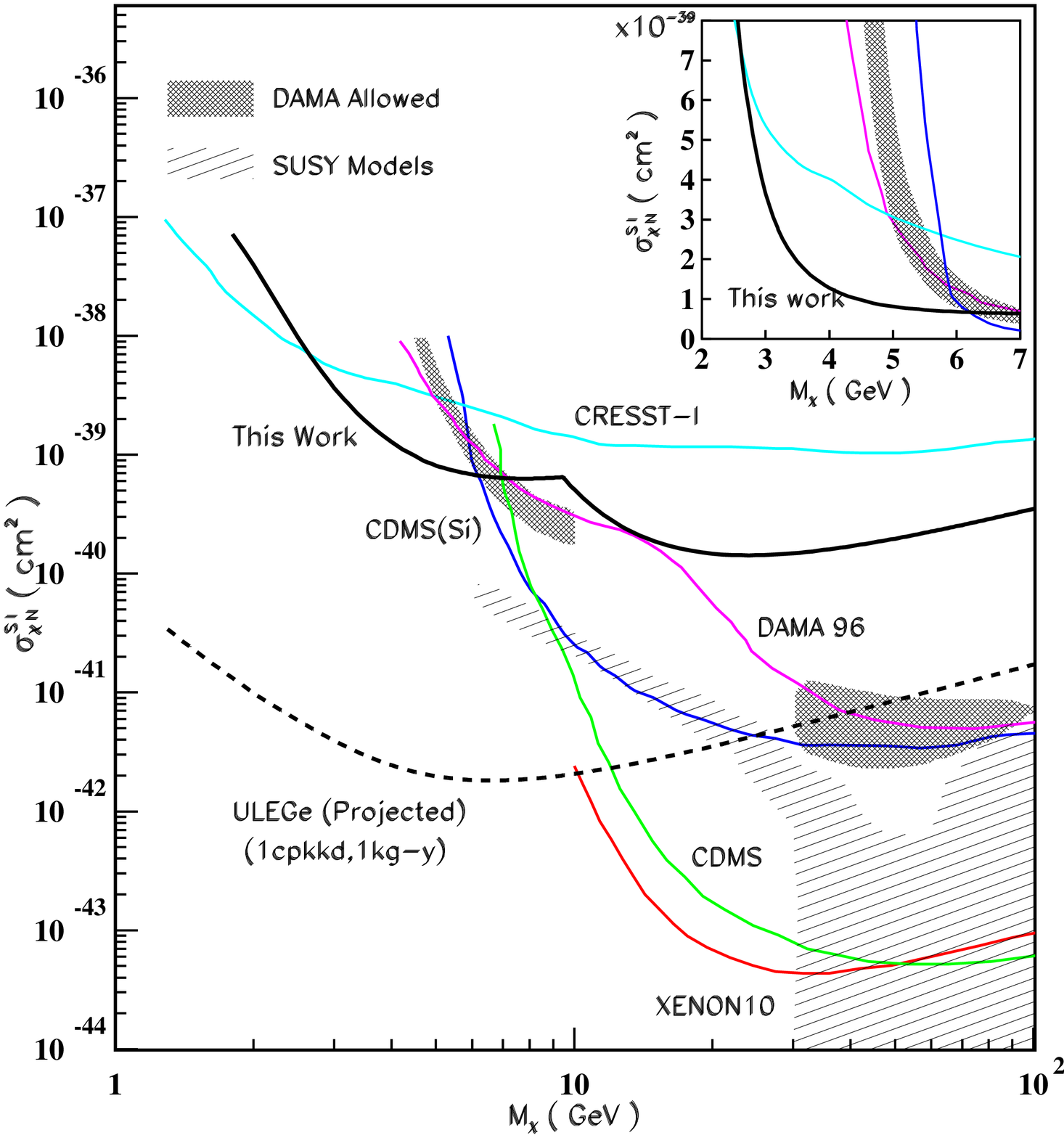}
\caption{
Exclusion plot of the
spin-independent $\chi$N
cross-section versus
WIMP-mass, displaying the
KS-ULEGe limits and
those defining the
current boundaries\cite{cresst1,cdmbounds}.
The DAMA-allowed regions\cite{damaallowed}
are superimposed.
The striped region is that favored
by SUSY models\cite{lightsusy}.
Projected sensitivities of full-scale
experiments are indicated as dotted lines.
}
\label{explotsindep}
\end{minipage}\hspace{2pc}%
\begin{minipage}{18pc}
\includegraphics[width=80mm]{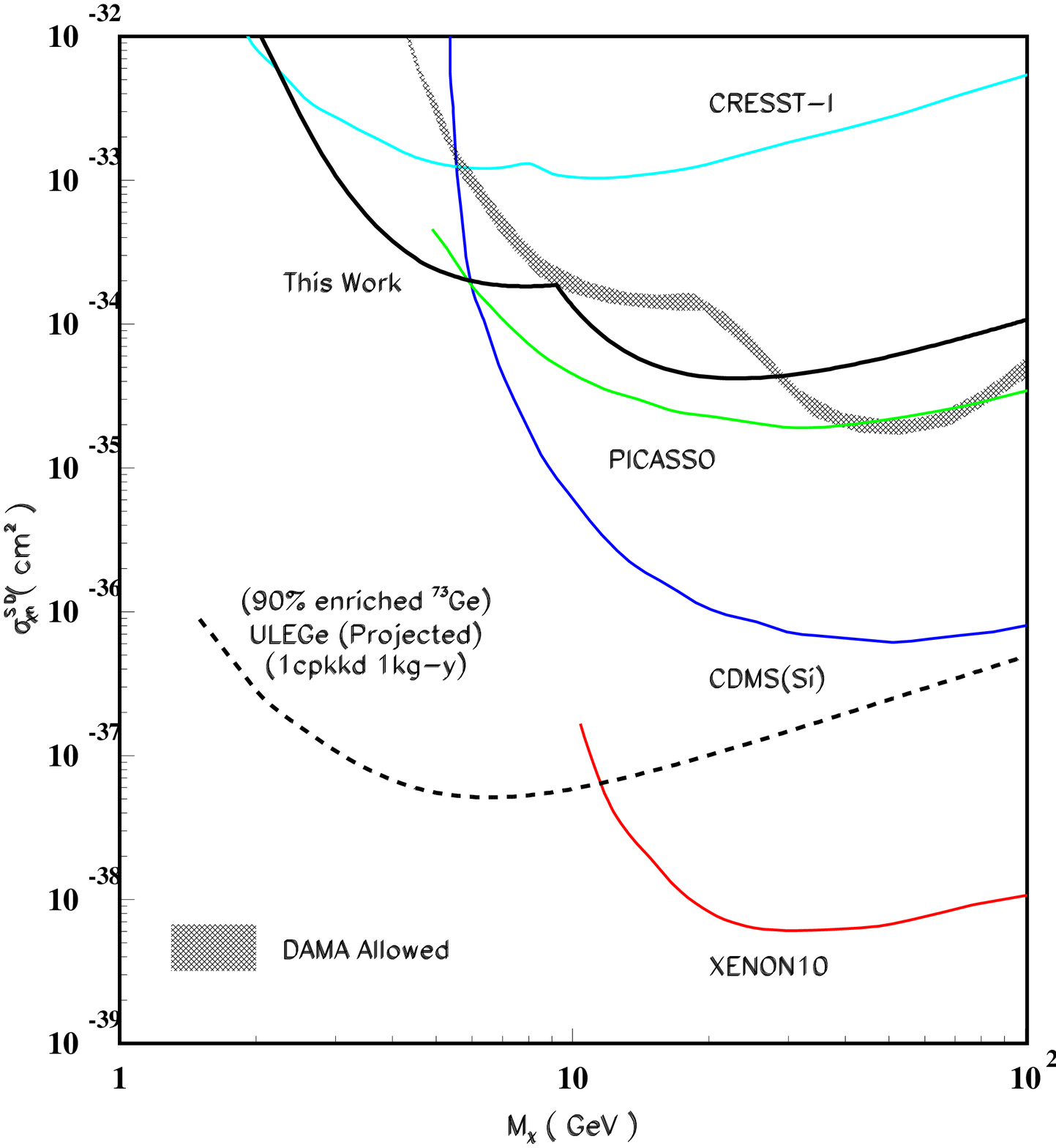}
\caption{
Exclusion plot of the
spin-dependent $\chi$-neutron 
cross-section versus
WIMP-mass. Same conventions
as those in Figure~\ref{explotsindep} are used.
}
\label{explotsdep}
\end{minipage}
\end{figure*}

Exclusion plots
on both $( \mwimp ,  \csnospin )$ 
and $( \mwimp ,  \csspin )$ planes
at 90\% confidence level
for galactically-bound WIMPs
were then derived, as
depicted in Figures~\ref{explotsindep}
and \ref{explotsdep}, respectively.
The DAMA-allowed regions\cite{damaallowed}
and the current exclusion boundaries\cite{cresst1,cdmbounds}
are displayed.
The ``model-independent'' approach
of Refs.~\cite{sdeptovey}
were adopted to extract limits on the
spin-dependent cross-sections.
Consistent results were obtained when
different $^{73}$Ge nuclear physics 
matrix elements\cite{ge73np} were adopted
as input.
The parameter space probed by the $^{73}$Ge in ULEGe
is complementary
to that of the CRESST-I experiment\cite{cresst1}
where the $^{27}$Al target is made up
of an unpaired proton instead.
New limits were set by
the KS-ULEGe data in both
$\csnospin$ and $\csspin$
for $\mwimp$$\sim$3$-$6 ~GeV.
The remaining DAMA low-$\mwimp$ allowed regions
in both interactions were
probed and excluded.
The observable nuclear recoils
at $\mwimp$=5~GeV and
$\rm\csnospin$=${\rm 0.5 \times 10^{-39} ~ cm^2}$(allowed)
and ${\rm 1.5 \times 10^{-39} ~ cm^2}$(excluded)
are superimposed with the measured spectrum
in the inset of Figure~\ref{bkgspect} for illustrations.
It is expected that
recent data from the COUPP\cite{coupp08} experiment 
can place further constraints in the 
spin-dependent plots of 
Figures~\ref{explotsdep}.

This work extends the bounds on WIMPs 
by making measurements 
in a new observable window of 100~eV$-$1~keV
in a low-background environment.
Understanding and suppression
of background at this sub-keV region is
crucial for further improvement.
There are recent advances in 
``Point-Contact'' Ge detector\cite{chicago} 
which offer potentials of scaling-up the detector
mass to the kg-range.
The mass-normalized external
background will be reduced in 
massive detectors due to self-attenuation\cite{cospa08}.
The potential reach of full-scale experiments with
1~kg-year of data and a benchmark background level of 
1~cpkkd is illustrated in Figures~\ref{explotsindep}
and \ref{explotsdep}.
Such experimental programs 
are complementary
to the many current efforts on 
CDM direct searches. 


\end{document}